 \documentclass[final,5p,times,twocolumn]{elsarticle}

\usepackage{color}
\usepackage{graphicx}
\usepackage{natbib}
\usepackage{bm}
\usepackage{amsmath}

\usepackage[nodots,nocompress]{numcompress}
\biboptions{super}
\journal{Physica D}
\date{}

\newcommand{\bfa}{{\bf a}}
\newcommand{\bfv}{{\bf v}}

\newcommand{\bfk}{{\bf k}}
\newcommand{\bfx}{{\bf x}}

\newcommand{\bfu}{{\bf u}}

\newcommand{\bfr}{{\bf r}}

\begin{document}
\begin{frontmatter}

  \title{Impact of trailing wake drag on the statistical
    properties and dynamics\\ of finite-sized particle in
    turbulence}

\author[lb1]{Enrico Calzavarini}
\cortext[cor1]{Corresponding authors:}
\ead{ecalzavarini@gmail.com}
\author[lb1]{Romain Volk}
\ead{romain.volk@ens-lyon.fr}
\author[lb1]{Emmanuel L\'ev\^eque}
\author[lb1]{Jean-Fran\c{c}ois Pinton}
\author[lb2]{Federico Toschi}
\address[lb1]{Laboratoire de Physique, Ecole Normale Sup\'erieure de Lyon,
CNRS and Universit\'e de Lyon, 46 All\'ee d'Italie, 69007 Lyon, France.\\
International Collaboration for Turbulence Research}
\address[lb2]{Dept.\,Physics and Dept.\,Mathematics \& Computer Science and J. M. Burgers Centre for Fluid Dynamics,\\
Eindhoven University of Technology, 5600 MB Eindhoven, The Netherlands.\\
International Collaboration for Turbulence Research.}

\begin{abstract}
  We study by means of an Eulerian-Lagrangian model the statistical
  properties of velocity and acceleration of a  neutrally-buoyant finite-sized particle
  in a turbulent flow statistically homogeneous and isotropic. The
  particle equation of motion, beside added mass and steady Stokes drag,
  keeps into account the unsteady Stokes drag force - known as
  Basset-Boussinesq history force - and the non-Stokesian drag based
  on Schiller-Naumann parametrization, together with the finite-size Fax\'en
  corrections.  We focus on the case of flow at low Taylor-Reynolds
  number, $Re_{\lambda} \simeq 31$, for which fully resolved numerical
  data which can be taken as a reference are available (Homann \& Bec
  {\bf 651} 81-91 {\it J. Fluid Mech.}
  (2010)).  
  Remarkably, we show that while drag forces have always minor effects
  on the acceleration statistics, their role is 
  important on the velocity behavior.
  We propose also that the scaling relations for the particle velocity
  variance as a function of its size, which have been first detected
  in fully resolved simulations, does not originate from
  inertial-scale properties of the background turbulent flow but it is likely to arise from
  the non-Stokesian component of the drag produced by the wake behind
  the particle. Furthermore, by means of comparison with fully resolved simulations, we show that the Fax\'en correction to the added mass has a dominant role in the particle acceleration statistics even for particle with size in the inertial range.
\end{abstract}



\end{frontmatter}
\section{Introduction}
The exact dynamics of a material particle in an inhomogeneous unsteady
flow involves nonlinear equations that can be treated analytically
only in approximate form \cite{maxey:1983,
  gatignol:1983,auton:1988,lovalenti:1993}. For this reason, several
simplified models for the hydrodynamic forces acting on a particle
have been proposed in the literature
\cite{michaelides:1997,loth:2009}. It is still unclear however to what
extent these models provide an accurate description in turbulent flow
conditions, even in an averaged or statistical sense.  A proper
statistical description of particle dynamics would be a first
important step towards building constitutive equations for the
particulate phase carried by turbulent fluids.  It would also be of
practical importance for the many environmental phenomena and
industrial applications in which particle suspensions in turbulence
are involved. We intend here to contribute to this goal by carrying out
refined simulations of particles in turbulent flow, and discriminate
whether the particle model employed leads to physically sound results,
in agreement with recent experiments and 
with fully resolved direct numerical simulations.

In previous studies, we have addressed the dynamics of small material
particles with a description based on a minimal Lagrangian model
accounting for pressure gradient, added mass term, and steady Stokes
drag force \cite{calzavarini:2008a,calzavarini:2008b}.  While this
system produces several features of particle dynamics, like
clustering and segregation as well as single-/multi-time statistics of
acceleration and velocity, it fails to predict some statistical properties,
particularly when the size of the particle is progressively increased
above the dissipative scale of turbulence \cite{volk:2008b}.  In order
to better understand these discrepancies we have focused on the case
of finite-sized and neutrally buoyant particles. We have proposed that
Fax\'en corrections are the essential ingredients to account for the
statistical properties of finite-sized particle acceleration in
turbulence \cite{calzavarini:2009}. Numerical predictions of the particle acceleration variance (and of its probability density function) based on the Fax\'en argument were compared with experiments made in wind-tunnel
\cite{qureshi:2007} and resulted in agreement with Von Karman flow
measurements \cite{brown:2009}. In a more recent work, experimental
measurements support also other trends highlighted by the Fax\'en model: the
effect of decrease of acceleration flatness as a function of the
particle-size and the corresponding growth of the correlation time of
the acceleration \cite{volk:2010}.\\
Another series of numerical studies were recently conducted by Homann
and Bec \cite{homann:2010} (HB in the following). These authors
employed a direct numerical approach. They tracked the motion of a
neutrally buoyant finite-sized particle in a turbulent flow by
enforcing the no-slip velocity at the particle
surface via a penalty method on the discretized Navier-Stokes
equation.  In such a way they have been able to access the dynamics of
a single finite-size particle in the diameter range $[2,16]\eta$ in a
moderately turbulent flow at $Re_{\lambda} =
32$. Both velocity and acceleration statistics were investigated.
Therefore, HB measurements provides a set of reference data against
which one can test particle Lagrangian models, as 
attempted here. The scope of this work however goes beyond the validation of a model
equation. We aim also at having a physical picture of the statistical
dynamics of particles.  We have specific questions in mind: What is
the statistical effect of the drag, particularly the trailing wake
drag, on the dynamics of a neutrally-buoyant finite-size particle in
turbulent flow?  Does it modify the acceleration statistics or rather
the velocity one? Is the role of Fax\'en correction still relevant for particles with size in the inertial range?
We will see how our study provides an answer to these questions and a possible interpretation of
 the phenomenological picture.\\
The paper is organized as follow. In section 2 we describe the approach adopted in this study, we introduce the Lagrangian modeling of the particle dynamics, its numerical implementation, and some expected trends for the particle velocity and acceleration in the vanishing-size limit.
In section 3 we present the results of the numerical study, starting from the particle Reynolds number behavior, and addressing then acceleration and velocity statistics as a function of the particle size. Comparison with direct numerical simulation data is  discussed in detail in section 4. In the conclusions we summarize the main results and and give suggestions for possible future  experimental/numerical investigations.

\section{Methods}
\subsection{Particle equation of motion}
We consider a Lagrangian equation of motion one-way coupled to a
continuum flow $\bfu \equiv \bfu(\bfx,t)$. Such an equation keeps into
account the pressure gradient and added mass term $(\sim D\bfu/Dt)$, the
drag force and the volume and surface Fax\'en corrections.  The drag
force is divided into three parts: the steady Stokes drag, the
unsteady Stokes drag force or History force, and the non-Stokesian
drag force.  All together it reads as follow:

\begin{eqnarray}
  \frac{d \bfv}{dt} &=&  \beta \left[ \frac{D\bfu}{Dt} \right]_{V} + \frac{3 \nu \beta}{r_p^{\ 2}} \left(  \left[ \bfu \right]_{S} - \bfv \right) \label{eq:1}\\
  &+& \frac{3 \beta}{r_p} \int_{t-t_h}^{t}  \left( \frac{\nu}{\pi(t-\tau)} \right)^{1 \over 2}\frac{d}{d \tau}  \left(  \left[ \bfu \right]_{S} - \bfv \right) d \tau  \label{eq:2}\\
  &+& c_{Re_p} \frac{3 \nu \beta}{r_p^{\ 2}} \left(  \left[ \bfu \right]_{S} - \bfv \right) \label{eq:3}
\end{eqnarray}

\noindent
where $r_p$ is the particle radius, $\nu$ the kinematic viscosity,
$\beta $ the density coefficient $\beta \equiv 3\ \rho_f / (\rho_f +
2\ \rho_p)$.  Following\cite{gatignol:1983} the Fax\'en corrections
are expressed as volume and surface average of the continuum fields
$D\bfu/Dt$ and $\bfu$ over a sphere of radius $r_p$ centered at the
particle position, respectively:
\begin{eqnarray}
  \left[ \frac{D\bfu}{Dt} \right]_{V} &=& (4/3 \ \pi r_p^3)^{-1} \int_V \frac{D\bfu}{Dt}(\bfx,t)\  d^3 x \\
  \left[ \bfu \right]_{S} &=& (4 \pi r_p^2)^{-1} \int_S \bfu(\bfx,t) \ d^2 x  
\end{eqnarray}
The history force is based here on the Basset-Boussinesq diffusive
kernel, $\sim(t-\tau)^{-1/2}$, while $t_h$ is the time over which the
memory effect is significant.
The non-Stokesian drag coefficient $c_{Re_p}$ models the effect of the
drag induced by the presence of a wake behind the particle. Of course
in a Lagrangian model of particle dynamics, which is only one-way
coupled to the fluid flow, no wake can be produced. Therefore, we
resort to a model: The well known Schiller-Naumann (SN) parametrization
\cite{schiller:1933}. The $c_{Re_p}$ coefficient, which is a function
of the particle-Reynolds number based on the diameter size
$d_p \equiv 2 r_p$ and on an estimator of the slip velocity, $Re_p \equiv \left| \left[ \bfu
  \right]_{S} - \bfv \right| d_p / \nu$, is chosen
to have the form $c_{Re_p} = 0.15 \cdot Re_p^{0.687}$ considered to be
a good approximation whenever $Re_{p} < 1000$
\cite{clift:1978}. We note also that direct numerical simulations of the
flow around a solid particle maintained fixed in a turbulent flow
shows a good agreement between the real force acting on the particle
(as computed from strain tensor at the surface of the particle) and
the drag computed from the slip velocity with Schiller-Naumann parametrization \cite{burton:2005,elghobashi:1992}.

\subsection{Numerical implementation}
We aim at studying the statistical signature of the different forces
acting on the particle.  For this reason in our numerical simulations
we follow the trajectories of four species (or families) of particles
with slightly different evolution equations. The first family is
described by (\ref{eq:1}), it includes only the added-mass term and
the steady Stokes drag and their Fax\'en corrections. It will be
called Fax\'en model with Stokes drag. The second family is defined by
(\ref{eq:1})+(\ref{eq:2}), hence it includes also the history
force. The third is based on (\ref{eq:1})+(\ref{eq:3}), therefore the
Schiller-Naumann correction is here included but not the history
force. Finally the forth family (\ref{eq:1})+(\ref{eq:2})+(\ref{eq:3})
keeps into account all the effects.
\begin{table}[!h]
  \begin{center}
    \begin{tabular}{ | l|  l| }
     \hline
	Particle type  & Eq.  \\  
	\hline
	 Fax\'en model with Stokes drag & (\ref{eq:1})  \\
	 $+$ History force &  (\ref{eq:1}, \ref{eq:2})\\	
	$+$ Non-Stokesian drag & (\ref{eq:1}, \ref{eq:3}) \\
	 $+$ Non-Stokesian drag $+$ History force &  (\ref{eq:1}, \ref{eq:2}, \ref{eq:3})\\
    \hline
    \end{tabular}
    \caption{\label{table:p}The four particle families considered in
      the present study and their respective
      equations of motion.}
  \end{center}
In total we integrate simultaneously the dynamics of $N_p =1.28\cdot 10^6$ particles. 
This ensemble is divided into 4 families, each family having  8 different sizes, in the diameter range $d_p \in [0,32]\eta$. This amount to $3.2\cdot 10^4$ particles per type.  
\end{table}

\subsubsection{Fax\'en Forces}
\begin{figure}
  \begin{center}
    \includegraphics[width=1.00\columnwidth]{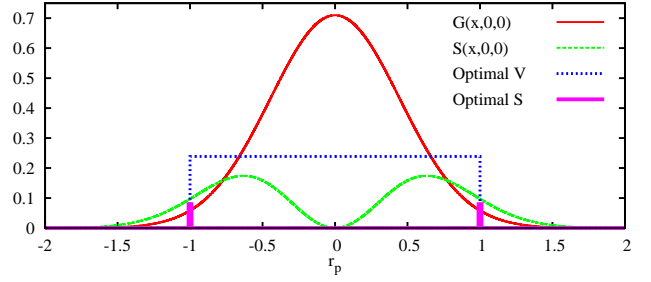}
    \caption{Real space one-dimensional projection, on the direction
      (x,0,0), of filter functions.  The volume gaussian filter is
      $G(\bfx) = (1/(\sqrt{2 \pi} \sigma))^3
      \exp(-\bfx^2/(2\sigma^2))$, while the surface convolution kernel
      turns out to be $S(\bfx) = (\bfx^2/(2\sigma^2))G(\bfx)$. The
      optimal shape of volume and surface filter functions are also
      shown. The volume filter function is a three dimensional spherical gate function with total volume $4/3, \pi r_p^3$, while the surface one is a delta function over the surface of the sphere but normalized in such a way that its volume integral is $4\pi r_p^2$.}
    \label{figure:filt}
  \end{center}
\end{figure}

The implementation of Fax\'en averages in our simulation is based on the Gaussian
approximation proposed in a previous study \cite{calzavarini:2009}.
The volume average of fluid acceleration at particle position is
replaced by a local interpolation at the particle position of the
continuum field after convolution by a three-dimensional Gaussian
envelope $G(\bfx)$, with unit volume and standard deviation
$\sigma$. Convolutions are efficiently computed in spectral space, the
volume averaged field hence reads:
\begin{eqnarray}
 \left[ \frac{D\bfu}{Dt}(\bfx, t) \right]_{V}  &\simeq&  \int_{L^3} G(\bfx')\  \frac{D \bfu}{Dt}(\bfx- \bfx', t) \, \, d^3x'\\ 
  &=&  \mathcal{ DFT}^{\ -1} \left[  \tilde{G}(\bfk)\  \frac{\tilde{D \bfu}}{Dt}(\bfk, t) \right]
\end{eqnarray}
where $\mathcal{ DFT}^{\ -1}$ denotes a discrete inverse Fourier
transform on a cubic grid with $N^3$ nodes, while the over script
$\sim$ indicates a direct Fourier transform, $\tilde{G}(\bfk) =
\exp({-\sigma^2 \bfk^2}/2) $ being the Fourier transform of $G(\bfx)$.
We note that by setting the standard deviation $\sigma \equiv
r_p/\sqrt{5}$, in the limit of small radii one gets $\tilde{G}(\bfk)
\simeq 1 - (r_p^2/10) \bfk^2 + {\cal O}(r_p^4)$ which leads to the
correct first order Fax\'en correction in real space, i.e.,
 $\bfu + (r_p^2/10) \Delta\bfu+ {\cal
  O}(r_p^4)$.  Analogously the surface average reads:
\begin{eqnarray}
  \left[ \bfu(\bfx, t)  \right]_{S} &=& \frac{1}{3 r_p^2} \frac{d}{dr_p} \left(  r_p^3  \  \left[ \bfu(\bfx, t)  \right]_{V}    \right)\\
  &=&  \mathcal{ DFT}^{\ -1} \left[ \left( 1 - \sigma^2 \bfk^2 / 3 \right) \tilde{G}(\bfk)\  \tilde{\bfu}(\bfk, t)  \right]. 
\end{eqnarray}
For clarity in Figure \ref{figure:filt} the shape of the two
convolution kernels (volume and surface) in real space are shown.
The figure also shows for comparison the so called \textit{optimal} convolution kernels, corresponding respectively to 
to a three dimensional spherical gate function of volume $4/3 \,\pi r_p^3$ for volume average and  to a delta function over a spherical shell
for surface average. Of course the implementation of such \textit{optimal} averages in real space would be computationally more expensive. In section \ref{sec:diss} we will investigate in detail the bias induced on the particle dynamics by the use of such a Gaussian approximation for Fax\'en corrections instead of the rigorous definition.

\subsubsection{History Force}
The Basset-Boussinesq history force can be computationally very
expensive. This is due to the fact that the integral which is 
involved should be performed at each time-step on the full particle
history.  Furthermore the diffusive kernel, $\sim(t-\tau)^{-1/2}$ has a
very slow decay and require a long memory time - virtually $t_h = \infty$ - to reach convergence. 
It is known however that the diffusive kernel overestimate the history force for particles
characterized by finite Reynolds numbers $Re_p$ \cite{mei:1992}.
The formation of a trailing wake either stationary, unstationary or even turbulent is always 
associated to history kernel which decay faster than the Basset-Boussinesq \cite{mordant:2000}.
Equivalently we can say that in $Re_p \gg 1$ conditions the Basset-Boussinesq history force should have a shorter memory time windows $t_{h}$. This latter idea have been exploited in the computational approach called windows method \cite{dorgan:2007}. Recently a more performant method based on the fit of the diffusive kernel tail via a series of exponential functions has also been proposed \cite{hinsberg:2010}.
In this study we adopt a simple windows approximation: instead of setting $t_h = \infty$ we
chose $t_h \simeq 10 \tau_{\eta}$ but keep the diffusive kernel functional form. This
choice is based on the observation that in the turbulent conditions considered
in our study after a time $10 \tau_{\eta}$ the Lagrangian signal $d (\left[ \bfu
\right]_{S} - \bfv ) /dt$ is already completely uncorrelated.  The
short memory on the history force is therefore not given by the
specific kernel form (which is indeed almost flat in our case) but
from the relatively short correlation time of the turbulent flow. With
this choice, $t_h $ corresponds approximately to $10^3$ time-steps of
our simulations, which are stored and used for the discrete estimation
of the history integral at each time-step. We note that our $t_h$ satisfy the criterion given in  ref. \cite{dorgan:2007,loth:2009}  for the windows method and it has a double extension in time steps respect to the time windows considered in the numerical validations considered in ref.\cite{hinsberg:2010}. We have also performed \textit{a posteriori} check in which the  pre-recorded $d (\left[ \bfu \right]_{S} - \bfv ) /dt$ signal has been used to compute the history force with different values of the windows length. 
This test has further confirmed the convergence and reliability of the adopted implementation.

\subsubsection{Eulerian dynamics}
A suitable turbulent flow is generated by integrating the
Navier-Stokes equation in a cubic box of size $L=2\pi$ with periodic
boundary conditions.  The flow is forced on the largest shells in
spectral space, on the wave-vectors for which the condition $\bfk^2
\leq 2^2 (2 \pi L)^2$ is satisfied. The force we adopt in this study
keeps fixed the amplitude of kinetic energy of the large
scales. More details concerning the values of relevant input and output quantities of the numerical simulation of this turbulent flow are provided in Table \ref{table}.

\subsection{Fax\'en corrections and small particle limit predictions}
In the study of Homann and Bec \cite{homann:2010} a derivation of the
functional behavior of the variance of particle velocity and
acceleration in the limit of vanishing particle diameters $d_p$ has
been proposed. The argument is based on a perturbative expansion of
the Fax\'en correction for the velocity. This reads as \footnote{Note
  that (\ref{eq:taylor}) corrects a typo contained in ref.
  \cite{homann:2010} on the numerical coefficient in front of $d_p^2
  \Delta \bfu$, which had affected all the predictions proposed in
  that study. Our calculations for eq. (\ref{eq:v}) provides a
  coefficient $5/3$, instead of $1/100$ given in
  \cite{homann:2010}. Furthermore in eq. (\ref{eq:a}) we find the coefficient
  $1/12$, and not $1/20$.}:
\begin{equation}\label{eq:taylor}
  \bfv \simeq \left[ \bfu \right]_{S}  \simeq \bfu + \frac{d_p^2}{24} \Delta \bfu + \mathcal{O}(d_p^4).
\end{equation}
Furthermore, the hypothesis of a spatially homogeneous particle
distribution in the limit $d_p \to 0$ is made.  By squaring
Eqn.~(\ref{eq:taylor}), retaining only quadratic terms in $d_p$, and
averaging over the particle ensemble and in time, $\langle \ldots
\rangle$, one gets:
\begin{equation}\label{eq:v}
  \langle \bfv^2 \rangle \simeq \langle \bfu^2 \rangle -  \frac{d_p^2}{12} \langle \bfu \Delta \bfu \rangle=
  \langle \bfu^2 \rangle -  \frac{d_p^2}{12} \frac{\varepsilon}{\nu} = \langle \bfu^2 \rangle -  \frac{5}{3}\left( \frac{d_p}{2 \lambda} \right)^2, 
\end{equation}
where $\varepsilon \equiv (\nu/2)\langle ( \nabla \bfu + (\nabla
\bfu)^T)^2\rangle = \nu\ L^{-3} \int_{L^3} \bfu \Delta \bfu \ d^3x$ is the
mean energy dissipation rate and $\lambda \equiv (5 \nu \langle \bfu^2
\rangle/ \varepsilon)^{1/2}$ is the Taylor micro-scale.  If we instead
(i) derive with respect to time Eqn.~(\ref{eq:taylor}), (ii)
make the assumption $D/Dt \simeq d/dt $, and then (iii) square and
average the result, we obtain an approximate prediction for the
acceleration variance
\begin{equation}\label{eq:a}
  \langle \bfa^2 \rangle \simeq \left< \frac{D \bfu}{Dt}^2 \right> -  \frac{d_p^2}{12} \left< \left|\left|\frac{D(\nabla \bfu)}{Dt}\right|\right|^2 \right>.  
\end{equation}
We will see in the following to which extent these approximations can
be considered appropriate  to describe the particle
behavior.  We note that in the simulations we have direct
access to the values $\langle( \left[ \bfu \right]_{S})^2 \rangle$ and
$\langle( \left[ D\bfu/Dt \right]_{V})^2 \rangle$ which can be used
for comparison. Finally, it is also worth noting that the particle
Reynolds number $Re_p$ is proportional to $|\bfv - \left[ \bfu
\right]_{S}|$, this means that in the small-particle limit the leading
order is $\mathcal{O}(d_p^4)$, hence one expects $Re_p \sim d_p^5$.
\begin{table*}[!t]
  \begin{center}
    \begin{tabular}{ c c c c | c c c c c c c c c c c c c}
      \hline
      $N^3$ & $\delta x$ & $\delta t$ & 
      $\nu$ & $\varepsilon$ & $\eta$ & $\tau_{\eta}$& $u_\mathrm{rms}$ & $\lambda$ & 
      $T_E$ & $L_E$ &  $Re_{\lambda}$ & $t_{tot}$ \\
      \hline    
      $128^3$ & $4.9 \cdot10^{-2}$ & $3.2\cdot10^{-3}$& 
      $4.4 \cdot 10^{-2}$ &  $7.5 \cdot 10^{-1}$
      &$1.0 \cdot 10^{-1}$ & $2.4 \cdot 10^{-1}$ &$1.2$ & $1.1$ &2.9 & 3.5 & 31& 192\\
      \hline
    \end{tabular}
    \caption{\label{table} 
      Parameters of the numerical simulation: $N$ number of grid
      points per spatial direction; $\delta x = 2\pi/N$ and $\delta t$
      are the spatial and temporal discretization; $\nu$ is the value
      of kinematic viscosity; $\varepsilon$ the mean value of the
      energy dissipation rate.  $\eta = (\nu^3/\varepsilon)^{1/4} $
      and $\tau_{\eta} = (\nu/\varepsilon)^{1/2}$ are the Kolmogorov
      dissipative spatial and temporal scales, $u_{rms}= \left(
        \overline{ \langle u_i u_i \rangle}_V/3 \right)^{1/2}$ the
      single-component root-mean-square velocity, $\lambda = (15\ \nu\
      u_{rms}^2 / \varepsilon )^{1/2}$ the Taylor micro-scale, $T_E =
      (3/2)u_{rms}^2/\varepsilon$ and $L_E= u_{rms}\, T_E$ are the
      Eulerian large-eddy-turnover temporal and spatial scales;
      $Re_{\lambda} = u_{rms}\ \lambda /\ \nu$ the Taylor scale based
      Reynolds number. $t_{tot}$ is the total simulation time in
      statistically stationary conditions and the total time-span of
      particle trajectories.}
  \end{center}
\end{table*}

\section{Results}
\subsection{Particle Reynolds number}
\begin{figure}
  \begin{center}
    \includegraphics[width=1.00\columnwidth]{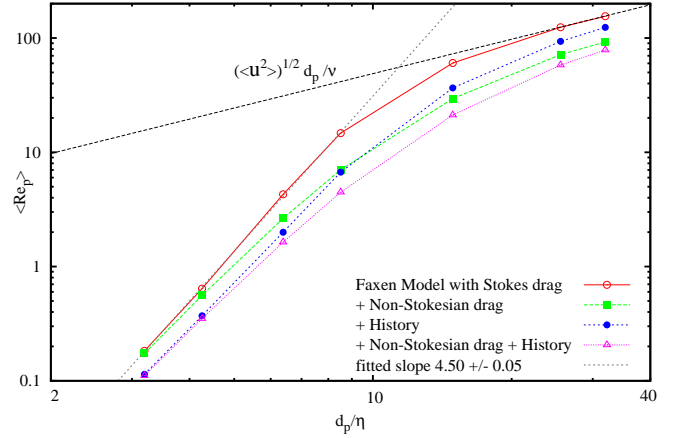}
    \caption{Mean particle Reynolds number, $\langle Re_{p} \rangle$,
      versus the diameter in Kolmogorov scale units, $d_p/\eta$, form the four different models.
      A  power-law fit  to the Fax\'en model with Stokes drag  on the smallest particle sizes is shown, we get a
      $4.5$ slope.  For the same model, the maximal Reynolds number $\bfu_{rms} d_p/\nu$ is reached by the largest
      particles.
      }
    \label{figure:rep}
  \end{center}
\end{figure}

We begin investigating the particle Reynolds number $Re_p$ as a
function of the particle diameter, with measurements reported in
Fig. \ref{figure:rep}. First we note that in the range $d_p \in [3.2,
32]\eta$ the mean Reynolds number varies considerably, three order of
magnitudes from $10^{-1}$ to about $10^2$.  We can immediately observe
that the particle models which are making use only of Stokes drag
forces - that is to say based on the assumption $Re_p < 1$ - can not
be considered entirely consistent. Such models underestimate the
actual drag on the particle. This is clearly noticeable for the
Fax\'en model with Stokes drag in the large-$d_p$ range, when $Re_p$
attains the maximal value $\bfu_{rms} d_p / \nu$, corresponding to a
ballistic particle velocity $\bfv$ not varying in time and not
correlated to the local fluid velocity $\left[ \bfu \right]_{S}$.
We note instead that for all the models in the small particle limit we
have a steeper scaling (slope $4.50 \pm 0.05$), close to the expected
$d_p^{\ 5}$. Hence, in the small particle regime $\bfv$ and $\left[ \bfu \right]_{S}$ 
are highly correlated, differing only by ${\cal
  O}(d_p^4)$ terms. We also notice that, while the history force
produces just a shift, the non-Stokesian drag term changes the slope
in the large particle regime.  This apparently minimal variations
have, as we will see later on, important consequences on the
statistics of the particle velocity variance.
\begin{figure}
  \begin{center}
    \includegraphics[width=1.00\columnwidth]{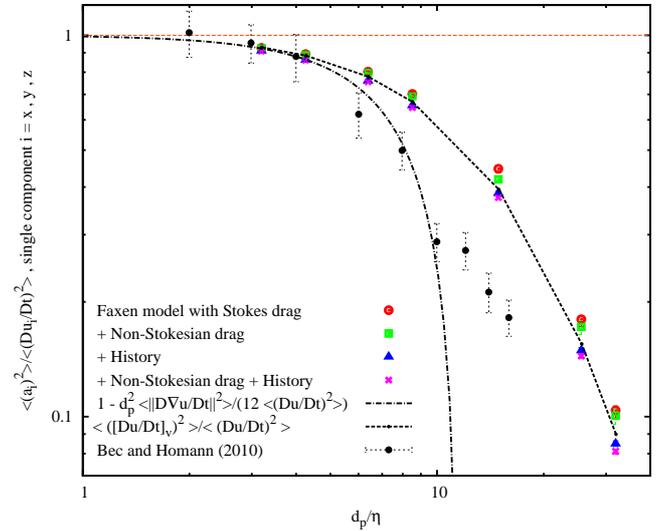}
    \caption{Normalized single-component particle acceleration
      variance for particles of different sizes: $\langle
      (a_i)^2 \rangle/\langle (D_t u_i)^2 \rangle$ vs. $d_p/\eta$. Here
      $\langle (D_t u_i)^2 \rangle$ is the fluid tracer acceleration
      variance or equivalently the Eulerian acceleration averaged over
      time and space. The behavior of the four different models adopted is
      reported.  The dashed-dotted line represent the behavior of particle acceleration variance
      expected to originate from Fax\'en corrections in the
      small particle limit (see eq (\ref{eq:a})). The dotted line represents the ratio
      of the variance of the volume filtered eulerian field $\left[ D\bfu/Dt \right]_V$
      to the fluid acceleration variance. 
      }
    \label{figure:a2}
  \end{center}
\end{figure}
\subsection{Acceleration statistics}
We examine now some statistical properties of the acceleration.
In Figure \ref{figure:a2} we show the behavior of the single-component
particle acceleration variance $\langle a_i^2 \rangle$ normalized by
the fluid acceleration variance $\langle (D_t u_i)^2 \rangle$ as a function of the particle size in
$d_p/\eta$ units. It is remarkable to note that all the particle
models leads to very similar results.  The History force or the
Non-Stokesian drag have no effect, at least for this observable. The
overall trend of the acceleration variance is dominated by the Fax\'en
Volume correction, $\langle a_i^2 \rangle \simeq \langle \left[
  D\bfu_i/D t \right]_V^2 \rangle$ with $\left[ D\bfu_i/D t \right]_V$
sampled homogeneously in space over the field (see
Fig. \ref{figure:a2}).  Equation (\ref{eq:a}) based on the first order
approximation, although qualitatively correct, fails to predict
quantitatively the measurements for $d_p > 4 \eta$.  In Figure
\ref{figure:a2} we have plotted the data points from HB
\cite{homann:2010}.  The agreement with our data is excellent up to
$d_p \simeq 4 \eta$, while the Lagrangian model shows a less
pronounced decrease (approximately by a factor of two) for larger
diameters. We will analyze the possible origin of these difference in section 4.
 
\begin{figure}
  \begin{center}
    \includegraphics[width=1.00\columnwidth]{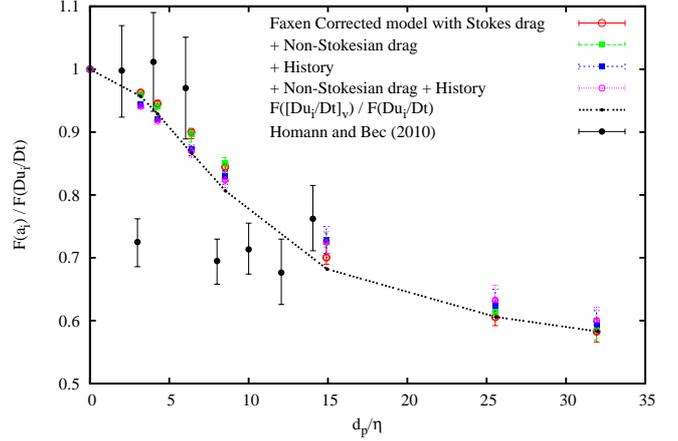}
    \caption{Single-component particle acceleration flatness,
      $F(a_i)$, normalized by the fluid tracer acceleration flatness,
      $F(D_t u_i)$, as a function of the particle diameter in $\eta$
      units. The dotted line represents the ratio
      of the flatness of the volume filtered eulerian field $\left[ D\bfu/Dt \right]_V$
      to the fluid acceleration flatness. }
    \label{figure:a3}
  \end{center}
\end{figure}

\begin{figure}
  \begin{center}
    \includegraphics[width=1.00\columnwidth]{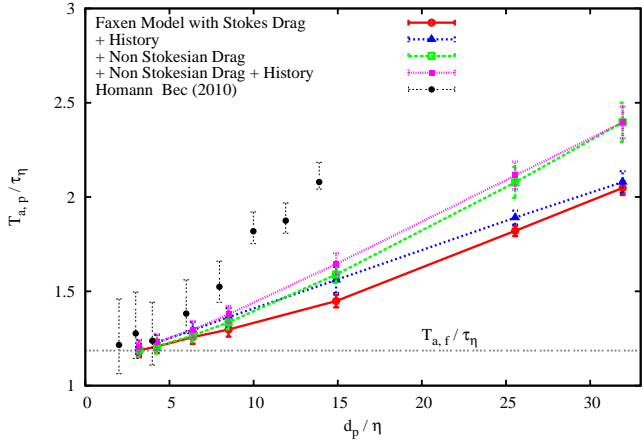}
    \caption{Correlation time of acceleration $T_{a, p}$, in
      $\tau_{\eta}$ units, as a function of the normalized particle
      diameter $d_p/\eta$.}
    \label{figure:ca}
  \end{center}
\end{figure}

In Figure \ref{figure:a3} we show the measurements of the acceleration
flatness $F(a_i) = \langle a_i^4\rangle/( \langle a_i^2\rangle)^2$
normalized by the fluid acceleration flatness $F(D\bfu_i/Dt)$ versus
particle size. As already noticed in \cite{calzavarini:2009}, and
experimentally verified in \cite{volk:2010} the flatness decreases with
increasing the particle size. Here the different Lagrangian models
lead only to small shifts in the flatness value, hence the picture
remains the same as for the variance.  Although, HB direct numerical simulations may suffers of lack convergence  the case of flatness, the measurements are in qualitative agreement with our simulations.
Finally we look at the correlation time of the particle
acceleration. As done before \cite{calzavarini:2009}, we define the
acceleration correlation time as the integral of the correlation
function from time zero till its first change of sign:
\begin{equation}
  T_{a, p} = \int_{0}^{\tau_0}C_{\textrm{a}_i}(\tau)\ d\tau \ ;  
  \quad  C_{\textrm{a}_i}(\tau) \equiv \frac{\langle \textrm{a}_i(t+\tau) \textrm{a}_i(t) \rangle}{\langle (\textrm{a}_i(t))^{2}\rangle} 
\end{equation}
where $C_{\textrm{a}_i}(\tau_0) = 0$.  We observe that, as soon as the
particle size grows, the acceleration correlation time $T_{a, p}$
deviates from the tracer value $T_{a, f} ( \simeq 1.2\ \tau_{\eta})$,
Fig. \ref{figure:ca}.  This growth, is a result of the Fax\'en
averaging (in fact it is absent when averaging is not included, see
the discussion in \cite{calzavarini:2009}) and comes from the fact
that in finite-sized particles Stokes drag becomes
negligible, leading to $d\bfv/dt \simeq \left[ D\bfu/Dt
\right]_V$.  We note however that there is also a significant
difference between the basic Fax\'en Stokes drag model, for which the
mechanism explained before is at work, and the model with
non-Stokesian drag, the latter one producing more correlation.  This
feature is rather surprising and has a different origin.  One may
think that including non-Stokesian drag, the effective response time
of the particle, i.e., $1/\tau_{eff} =(1+C_{Re_p})3 \nu \beta / r_p^2$
is reduced, therefore the acceleration should be correlated on a
shorter time scale $\sim \tau_{eff}$. As we will see later in section
\ref{sect:v} when non-Stokesian drag is active, the drag is never
negligible and $\bfv \simeq \left[ \bfu \right]_S$, hence for the
acceleration $d\bfv/dt \simeq d\left[\bfu \right]_{S}/dt$. It is clear
that $d\left[\bfu \right]_{S}/dt = \partial_t \left[\bfu \right]_{S} +
\left[\bfu \right]_{S}\cdot \partial \ \ \left[\bfu \right]_{S}$ does
not have the sub-grid (sub particle-size) correlations included in
$\left[ D\bfu/Dt \right]_V = \partial_t \left[\bfu \right]_{v}+
\left[\bfu\cdot{\bm\partial}\ \ \bfu \right]_{V}$, which are
correlated on shorter timescales.\\  Any Lagrangian model equation seems however 
to underestimate the real $T_{a, p}$ resulting form the HB direct numerical
simulations.  This fact has been also noticed, in a comparison of Fax\'en Lagrangian model with experimental data
\cite{volk:2010}.  

\subsection{Velocity statistics}\label{sect:v}
\begin{figure}
  \begin{center}
    \includegraphics[width=1.00\columnwidth]{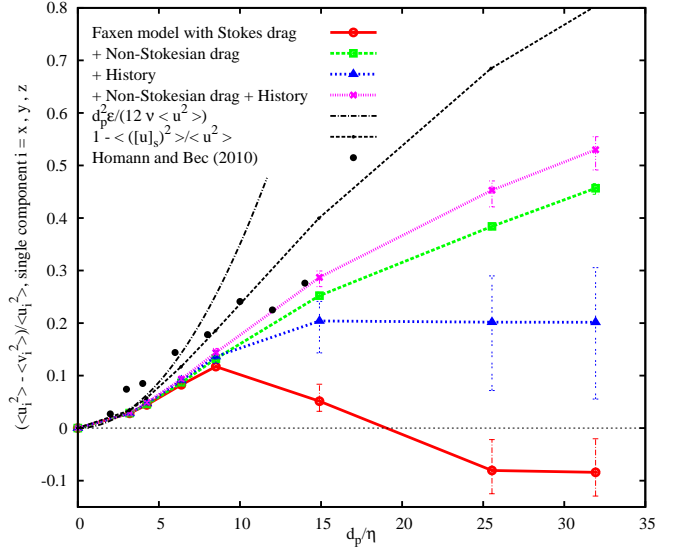}
    \caption{Deviation of the particle velocity variance from the
      fluid value, as a function of the dimensionless particle
      diameter $d_p/\eta$. The behavior of the four different models adopted is
      reported.  The dashed-dotted line represent the deviation
      from the fluid root-mean-square (r.m.s.) velocity
      that is expected to originate from Fax\'en corrections in the
      small particle limit (see eq (\ref{eq:v})). The dashed line represents the deviation
      of the variance of the surface filtered eulerian field $\left[ \bfu \right]_S$
      from the unfiltered velocity variance.}
    \label{figure:v2}
  \end{center}
\end{figure}

Contrary to the acceleration's case the
velocity particle statistics is deeply affected by the form of the
particle dynamical equation. In Figure \ref{figure:v2} we show as a function of the particle size the
measurements of the deviation of normalized single-component particle
velocity variance with respect to the velocity variance of the
turbulent flow:  $(\langle \bfu_i^2 \rangle - \langle \bfv_i^2
\rangle)/\langle \bfu_i^2 \rangle$.  The first model - based simply on
Fax\'en terms and Stokes drag - predicts for this quantity a non-monotonous
behavior, leading for the bigger particle sizes to a  velocity
variance even larger than the fluid one. This results is rather
unphysical (as it is not possible for a particle to be on average more
energetic then the flow by which it is driven and transported) and
clearly it represents a limitation of the basic Fax\'en model with
Stokes drag.  This limitation is readily cured whenever an extra drag
force is added. Among History and non-Stokesian drag it is definitely
the latter one bringing the most significant changes. The non-Stokesian drag
reduces the kinetic energy of the particle as compard to the one of a
fluid tracer, this energy decreases monotonically with the particle
size.  For comparison, in Figure \ref{figure:v2} we have also
reported  the prediction in the limit of
small-particles, eq. (\ref{eq:v}). As for the acceleration, this
analytical prediction seems to be a good approximation to the
measurements up to $d_p \simeq 4 \eta$.  We also report in Figure
\ref{figure:v2} the value of the variance of the Eulerian filtered
field $\left[ \bfu \right]_S$. One can note that by adding more drag
to the basic Stokes term, the particle velocity approaches the
filtered fluid velocity variance, i.e. $\bfv \to \left[ \bfu
\right]_S$.  We note that the prediction of the Lagrangian models
keeping into account all the considered effects
eqs. (\ref{eq:1})-(\ref{eq:3}), agree well with the HB data.
Apart from the parabolic ($\sim d_p^2$) behavior for vanishing
particle sizes no clear scaling of the normalized velocity variance
can be detected form our measurement.
In HB it was proposed that a scaling regime with slope $d_p^{2/3}$
would appear out of the Fax\'en dominated regime. This was attributed
to the effect of the background turbulent flow, via the assumption $\langle v_i^2 \rangle \sim \langle (\delta u_{d_p})^2\rangle \sim d_p^{2/3}$
where  the Kolmogorov scaling relation for the
second order eulerian structure function $\langle (\delta u_r)^2 \rangle \equiv \langle
(\bfu(\bfx+\bfr)-\bfu(\bfx))\cdot \hat{\bfr}^2 \rangle \sim r^{2/3}$ is implied.
Here we would like to advance another
explanation, the different scaling in that intermediate regime seems to be
an effect of the  drag term and in particular of the non-Stokesian drag (see again Figure \ref{figure:v2}).  We note that
non-Stokesian drag term included in our model equations based on SN parametrization accounts  rather for the effect of a stationary wake
behind the particle than for wake generated turbulent fluctuations. 
Given the good agreement of our data with HB measurements, our guess is that
the background turbulent fluctuations plays only a minor role in determining the particle velocity statistics.

\begin{figure}
  \begin{center}
   \includegraphics[width=0.85\columnwidth]{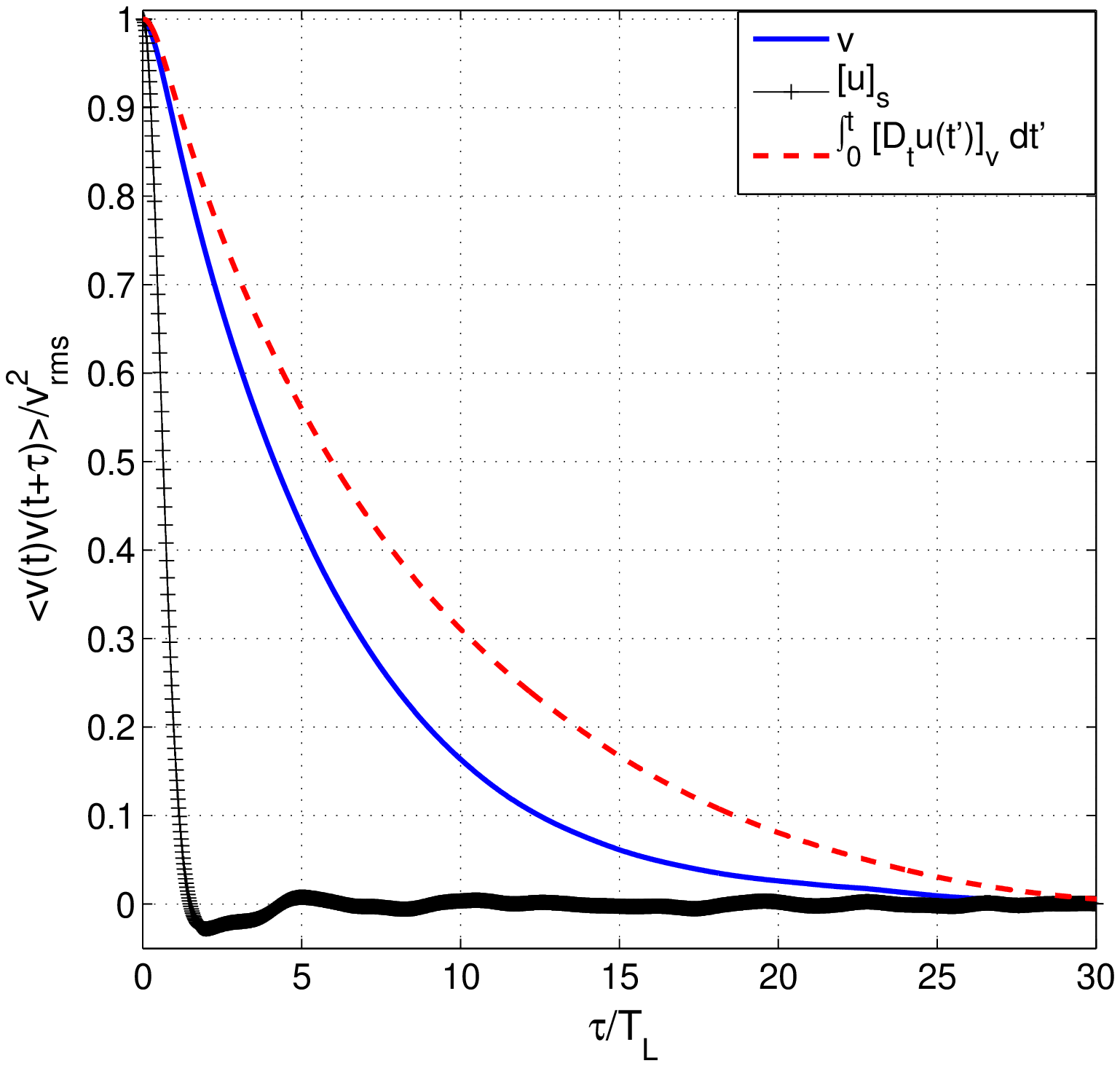}
	\setlength{\unitlength}{1.0cm}
	\begin{picture}(1,1)
	\put(-5.8,5.0){\small Fax\'en Model with Stokes drag}
	\end{picture}
    \includegraphics[width=0.85\columnwidth]{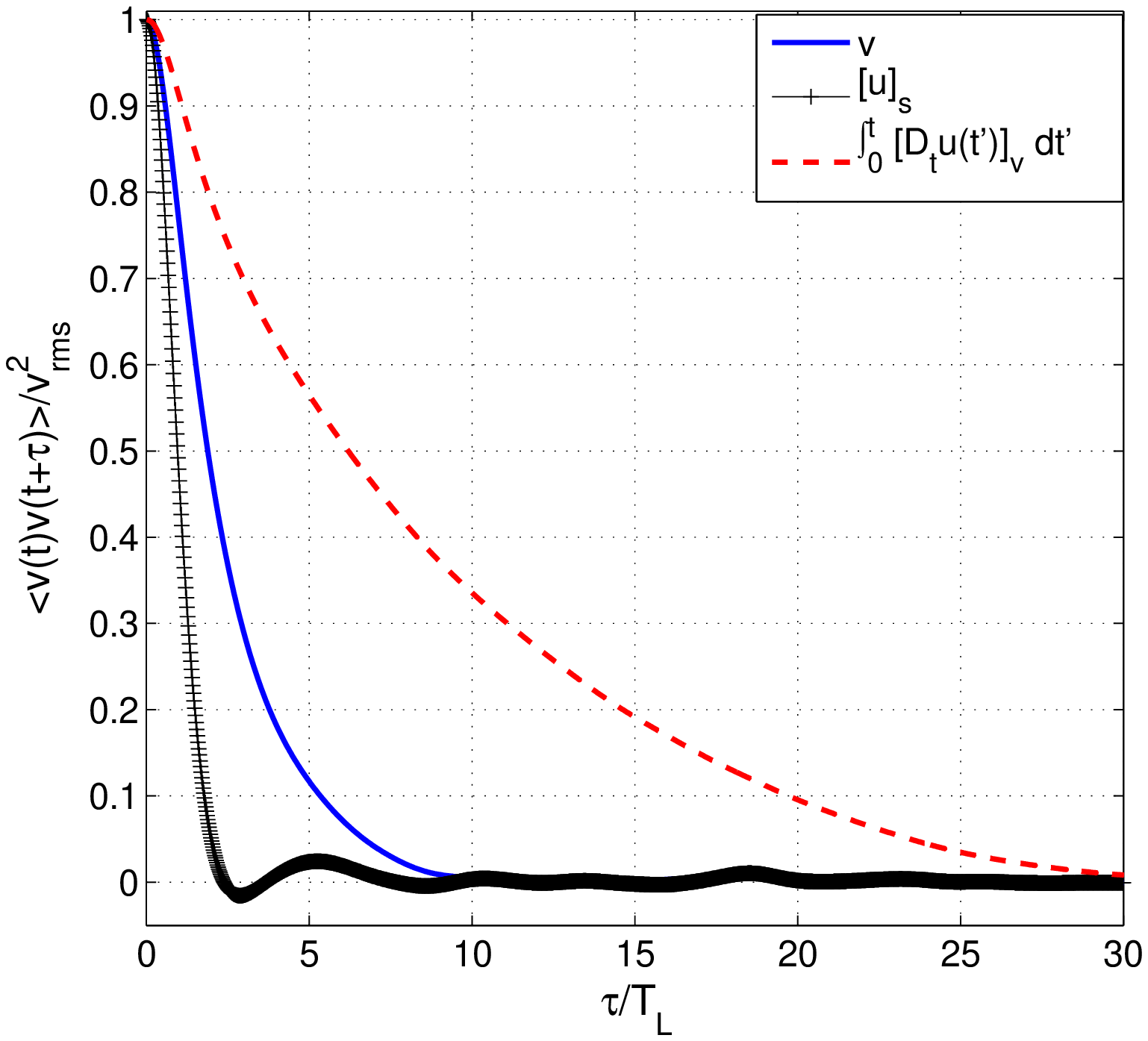}
    	\setlength{\unitlength}{1.0cm}
	\begin{picture}(1,1)
	\put(-6.0,5.5){\small + non-Stokesian drag}
	\end{picture}
    \caption{Correlation function of particle velocity $\bfv$, of the
      filtered fluid velocity along the particle trajectory $\left[
        \bfu \right]_S$, and of the time integral of fluid
      acceleration $\int_0^t \left[ D_t\bfu(t') \right]_V dt'$ for
      particle size $d_p = 32 \eta$. Fax\'en model with Stokes drag
      (top), with non-Stokesian drag (bottom).}
    \label{figure:cvv}
  \end{center}
\end{figure}

The non-Stokesian term has also important consequences on the velocity
correlation time.  Such a time can be defined as the time integral of
the correlation function:
\begin{equation}
  T_p = \int_{0}^{+\infty}C_{\textrm{v}_i}(\tau)\ d\tau \ ;  
  \quad  C_{\textrm{v}_i}(\tau) \equiv \frac{\langle \textrm{v}_i(t+\tau) \textrm{v}_i(t) \rangle}{\langle (\textrm{v}_i(t))^{2}\rangle} 
\end{equation}
As we have already mentioned, the Stokes drag force alone is not effective in
slowing down the velocity especially for large particles.  This
produces quasi-ballistic trajectories, governed by $\bfv(t) \simeq
\int_0^{t} \left[ D_t \bfu(t') \right]_V dt'$, that tend to have very
large correlation time. The non-Stokesian (or wake) drag provides instead a way to reduce
particle speed, $\bfv(t) \simeq \left[ \bfu \right]_S$ and its
correlation time. This is evident from Figure \ref{figure:cvv} where
the correlation functions for different terms for a large Stokesian
and a large non-Stokesian particle are compared.

\begin{figure}
  \begin{center}
    \includegraphics[width=1.00\columnwidth]{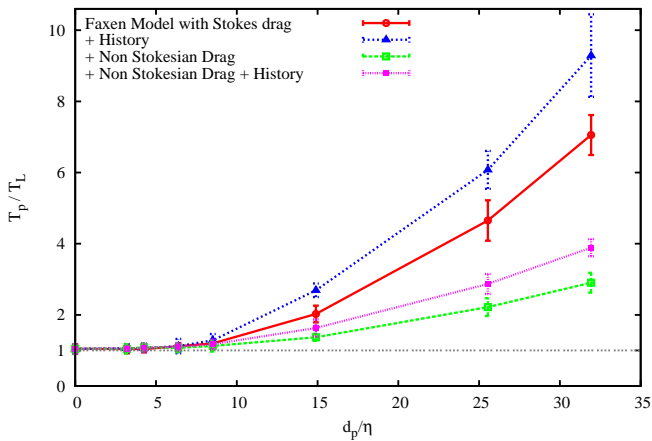}
    \caption{Integral correlation time of particle velocity $T_p$ as a
      function of the diameter $d_p$.  $T_p$ is made dimensionless by
      the integral velocity correlation time of a Lagrangian tracer
      $T_L$, while the diameter is normalized by the dissipative scale
      $\eta$.}
    \label{figure:tv}
  \end{center}
\end{figure}

On Figure \ref{figure:tv} the trend of the velocity correlation time
as a function of the particle size is shown. Non-Stokesian drag
produces a reduction of $T_p$ of more than 100\% as compared to the
purely Stokesian case.
%
%
\section{Discussion}\label{sec:diss} 
In previous sections we have shown that the agreement of the Lagrangian model, even in its complete form eqs. (\ref{eq:1}),(\ref{eq:2}),(\ref{eq:3}), with the measurements from HB direct numerical simulations is rather satisfactory for the variance of the velocity as a function of the particle size. However, we observe systematic deviations when acceleration is concerned.  
Here we would like to discuss the causes of these discrepancies more in detail.
We can advance the following hypothesis:\\
i) The differences originate from a limitation of
the model which takes into account volume and surface averages only in the approximate form of a Gaussian
convolution. Although this approximation is well tuned for the first
order Fax\'en correction, it might be less accurate for larger particles, when roughly $d_p > 4\eta$.\\ 
ii) The model neglect the effect of particle interaction with its own wake. 
This effect, while negligible in a flow with a large mean component, might become relevant in the isotropic
flow conditions considered here (and in HB work) where a particle can cross a region previously perturbed by its own wake.\\
iii) The observed discrepancies may come from differences in the simulated
turbulent flows  - in fact differences in the forcing at such small
Reynolds number $Re_{\lambda} \sim 32$ may have consequences even on the small scale statistics.
\begin{figure}[t]
  \begin{center}
    \includegraphics[width=0.95\columnwidth]{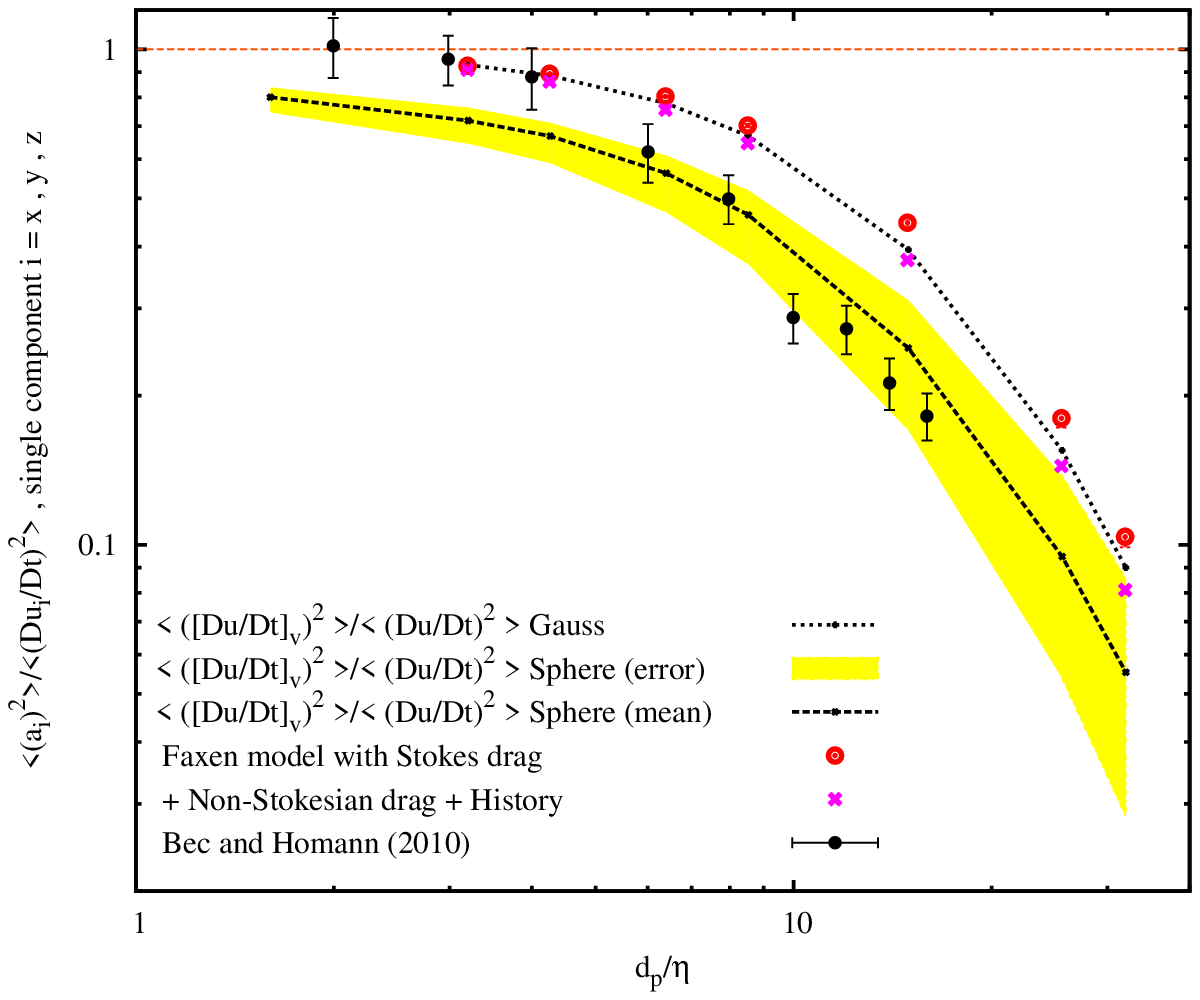}
    \includegraphics[width=0.95\columnwidth]{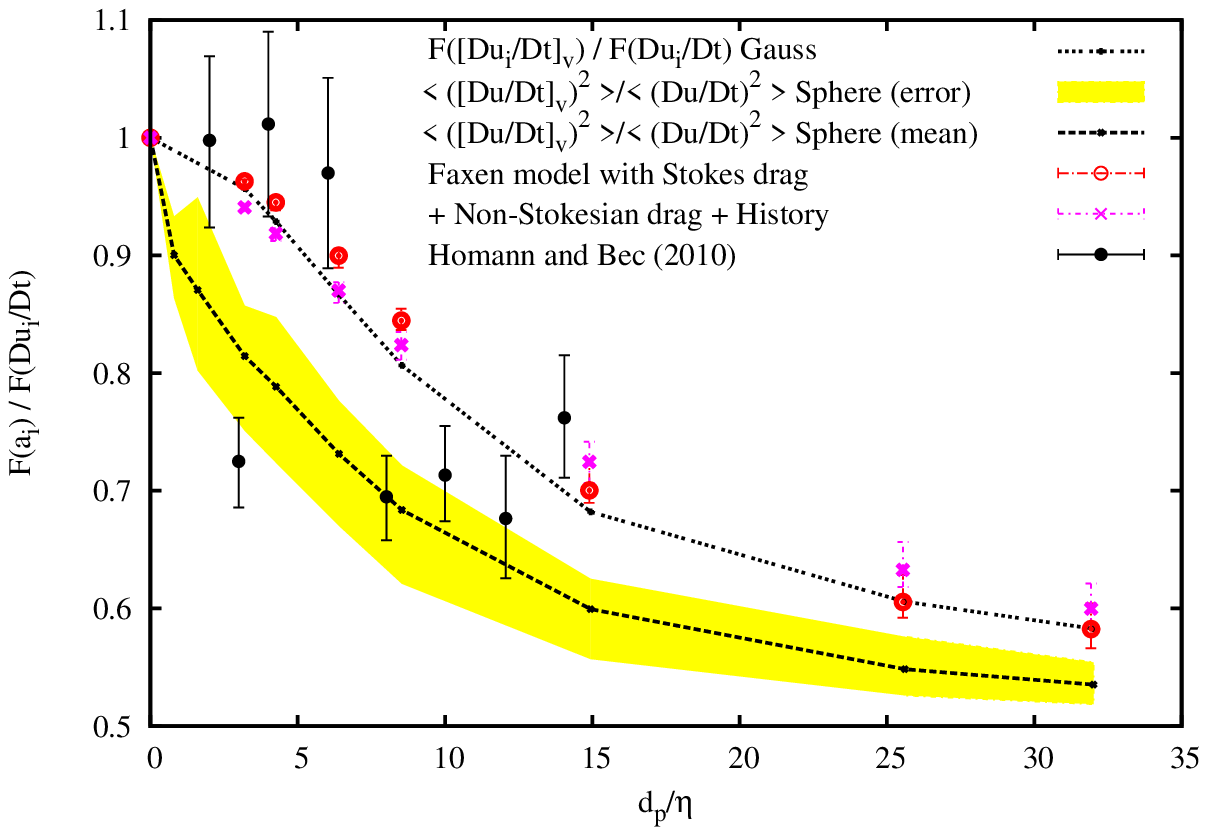}
    \caption{Comparison between acceleration variance (top panel) and  flatness (bottom panel)
We show data coming from Gaussian convolutions, same as in fig. \ref{figure:a2}, of the fluid acceleration field $D\bfu/Dt$ (dotted line), the quantity has been computed run-time therefore statistical errors are in this case of the order of the line thickness.
Averages over spherical volumes (dashed line) are affected by a larger statistical uncertainty (yellow shaded region) which comes from the differences in the measurements between the three cartesian components. Measurements from the Lagrangian model based on eq (\ref{eq:1}) and eq. (\ref{eq:1}),(\ref{eq:2}),(\ref{eq:3})  are reported, together with HB data.}
    \label{figure:sphere}
  \end{center}
\end{figure}
The point i) can be investigated carefully. In order to see if there is any bias induced by Gaussian averaging as compared to the mean over a sphere, we have averaged the a field $D\bfu(\bfx,t)/Dt$ over a large number of spheres of diameters  $d_p$,  uniformly distributed over random locations in space.
This procedure is repeated over 250 Eulerian $D\bfu/Dt$ snapshots, equally spaced in time over an interval of $20\ T_E$.
Although this method is not precise for small sphere diameters when only few points of the discretized $D\bfu/Dt$  field enter into the average, for large diameters the average converges rapidly to the correct (continuum-limit) values.
In figure \ref{figure:sphere} we report the results of these measurements for the variance and the flatness of $[D\bfu/Dt]_{V}$(sphere), as a function of $d_p$ and we compare it with the Gaussian averages $[D\bfu/Dt]_{V}$ (Gauss), and with the measurements on the Lagrangian models and  DNS data.
The result shows that the sphere average is sensibly different from the Gaussian convolution. For small particles the discretization bias fails to match the analytical prediction (\ref{eq:v}), which is instead well captured by the Gaussian distribution. For larger particle sizes the sphere average shows a stronger decrease of the variance as compared to the Gaussian case. Remarkably in this region the average over spheres agrees with the HB results. A similar scenario appears for the flatness fig. \ref{figure:sphere} (bottom). 
This finding is important at least for two reasons. First it shows that the Fax\'en correction to the fluid acceleration plays a central role in the particle acceleration statistics not only for a particle in the upper dissipative range $d_p<4 \eta$ but also for a particle with size is in the inertial range $d_p  \leq 32 \eta$.
Second, it shows that, although the Gaussian convolution approximation is an efficient method to solve particle dynamics, it has limitations which becomes important approximately when $d_p > 4 \eta$.  
The above observations have also an impact on the time-statistics: since a reduction of the variance of the filtered acceleration 
is associated to a slower fluctuations in time, we expect the correlation time of acceleration to increase for finite-sized particles. Hence the picture in fig \ref{figure:ca} would change, solving the observed mismatch towards HB simulations and experiments\cite{volk:2010}. However, this latter point may find a confirmations only in further studies.\\
Evaluating the impact of point ii) is unfortunately not possible in the framework of the present model. One would need to introduce a coupling (so called two-way coupling) between the particle and the fluid enforcing conservation of the total momentum.\\
Finally on point iii), for completeness we shall note that there are
some differences between the turbulent flows simulated by Homann \&
Bec and the one used here. While we adopt a forcing which keeps
constant the energy on the largest Fourier modes (amplitude-driving),
HB also uses a random forcing of the phase in Fourier space
(phase-driving). While these differences have no effect in fully
developed turbulent flows, at the small Reynolds number considered
here, $Re_{\lambda}\simeq 32$, they might have an impact. We see
indeed that while we find $\langle a_i^2 \rangle \varepsilon^{-3/2}
\nu^{1/2} = 1.1$, $F(a_i)= 5.7$, Homann \& Bec report 
$ \langle a_i^2 \rangle \varepsilon^{-3/2}
\nu^{1/2}=1.3$ and $F(a_i)= 8.4$, hence a flow slightly more
intermittent at small-scales. 
This prevents us, for instance,  from a direct comparison on the shape of the probability density functions of acceleration and velocity.

\section{Conclusions}
In this study we have focused on the statistical properties of
acceleration and velocity of finite-sized neutrally-buoyant particles driven in a moderately
turbulent homogeneous and isotropic flow. We have adopted a Lagrangian
model particle equation which keeps into account inertia effect, size
effects, and the drag forces resulting both from a Stokes flow around the particle and form an asymmetric trailing wake state via
Schiller-Naumann modeling. We have studied the contribution of
these forces separately, in particular the drag force has been divided
into three components: Stokes, History, and non-Stokesian force.

We find that the drag forces have minor effect on the statistical
properties of acceleration. Acceleration statistics seems to be dominated by inertia effect and by the Fax\'en corrections, whose
influences extends also over particle with size in the inertial turbulent range.

On the opposite drag forces have important effects on the time integral of the acceleration, that is to say
on the velocity statistics. This is particularly evident when
considering the trend of the second order statistical moment (the
variance) as a function of the particle size. For the case of
neutrally buoyant particle analyzed here, the variance of the particle velocity from
the different Lagrangian models start to separate at $d_p>8 \eta$, corresponding
to a particle Reynolds number $Re_p \sim {\cal O}(10)$. Above this
threshold History and non-Stokesian drag have a dominant role.
This lead us to propose that the trend observed for the particle velocity
variance as a function of its size does not originate from
inertial-scale properties of the background turbulent flow but arise from
the non-Stokesian component of the drag produced by the wake behind
the particle.

The effects detected in the velocity statistics are relevant for
studies of particle dispersion in turbulence. For instance a simple finite-size particle model based on Stokes drag or history force only, would overestimate the average particle dispersion from a fixed source in space, and similarly for pair separation. 
Therefore, this study suggests that in order to validate Lagrangian models one should look not only into
the small scale acceleration statistics - as done up to now in many studies - but also into velocity and
possibly dispersion and pair separation statistics.
 
On the numerical side, we have shown that the Gaussian convolution approximation, despite its computational efficiency  
is not accurate when particles  much larger than $4 \eta$ are involved. On the other hand spherical averages in real space, as the one performed in our test, would be computationally very expensive and not enough accurate for small size particles (in particular it is not possible to capture the laplacian correction (\ref{eq:taylor}) when only few grid points are used).  Future numerical studies should find a trade off between efficient computations and accuracy. A possible way, which however needs careful scrutiny and tuning,  is to consider the implementation of convolution kernel functions with sharper boundaries. 
Furthermore, the effect of two-way coupling in this type of homogeneous and isotropic turbulent flow deserve to be studied.

\textit{Acknowledgments:} The authors are thankful to A. Naso and
M. Bourgoin for useful discussions.  Support from COST Action MP0806
``Particles in turbulence'', from ANR-BLAN-07-1-192604 is
acknowledged.  E.C. was partially supported by the HPC-Europa2
visiting program. Numerics for this study have been performed at SARA
(The Netherlands), CINECA (Bologna, Italy) and CASPUR (Roma, Italy).
Numerical data of particles trajectories are available on iCFDdatabase
({\tt http://mp0806.cineca.it/icfd.php}) kindly hosted by CINECA
(Italy). More informations available upon request to the authors.

\bibliographystyle{model3a-num-names} \bibliography{biblio_fsm}

\begin{thebibliography}{22}
\providecommand{\natexlab}[1]{#1}
\providecommand{\url}[1]{\texttt{#1}}
\providecommand{\urlprefix}{URL }
\expandafter\ifx\csname urlstyle\endcsname\relax
  \providecommand{\doi}[1]{doi:\discretionary{}{}{}#1}\else
  \providecommand{\doi}{doi:\discretionary{}{}{}\begingroup
  \urlstyle{rm}\Url}\fi
\providecommand{\eprint}[2][]{\url{#2}}
\providecommand{\bibinfo}[2]{#2}
\ifx\xfnm\undefined \def\xfnm[#1]{\unskip,\space#1}\fi
\makeatletter\def\@biblabel#1{#1.}\makeatother
\bibitem[{Maxey and Riley(1983)}]{maxey:1983}
\bibinfo{author}{Maxey\xfnm[ M.R.]}, \bibinfo{author}{Riley\xfnm[ J.J.]}.
\newblock \bibinfo{title}{Equation of motion for a small rigid sphere in a
  nonuniform flow}.
\newblock \emph{\bibinfo{journal}{Phys Fluids}}
  \bibinfo{year}{1983};\hspace{0pt}\textbf{\bibinfo{volume}{26}}(\bibinfo{numb%
er}{4}):\bibinfo{pages}{883--889}.
\bibitem[{Gatignol(1983)}]{gatignol:1983}
\bibinfo{author}{Gatignol\xfnm[ R.]}.
\newblock \bibinfo{title}{The fax\'en formulae for a rigid particle in an
  unsteady non-uniform stokes flow}.
\newblock \emph{\bibinfo{journal}{J Mecanique Theorique et Appliqu\'ee}}
  \bibinfo{year}{1983};\hspace{0pt}\textbf{\bibinfo{volume}{1}}(\bibinfo{numbe%
r}{2}):\bibinfo{pages}{143--160}.
\bibitem[{Auton et~al.(1988)Auton, Hunt and Prud'homme}]{auton:1988}
\bibinfo{author}{Auton\xfnm[ T.]}, \bibinfo{author}{Hunt\xfnm[ J.]},
  \bibinfo{author}{Prud'homme\xfnm[ M.]}.
\newblock \bibinfo{title}{The force exerted on a body in inviscid unsteady
  non-uniform rotational flow}.
\newblock \emph{\bibinfo{journal}{J Fluid Mech}}
  \bibinfo{year}{1988};\hspace{0pt}\textbf{\bibinfo{volume}{197}}:\bibinfo{pag%
es}{241--257}.
\bibitem[{Lovalenti and Brady(1993)}]{lovalenti:1993}
\bibinfo{author}{Lovalenti\xfnm[ P.M.]}, \bibinfo{author}{Brady\xfnm[ J.F.]}.
\newblock \bibinfo{title}{The hydrodynamic force on a rigid particle undergoing
  arbitrary time-dependent motion at small reynolds number}.
\newblock \emph{\bibinfo{journal}{J Fluid Mech}}
  \bibinfo{year}{1993};\hspace{0pt}\textbf{\bibinfo{volume}{545}}:\bibinfo{pag%
es}{561--605}.
\bibitem[{Michaelides(1997)}]{michaelides:1997}
\bibinfo{author}{Michaelides\xfnm[ E.E.]}.
\newblock \bibinfo{title}{Review - the transient equation of motion for
  particles, bubbles, and droplets}.
\newblock \emph{\bibinfo{journal}{J Fluid Eng}}
  \bibinfo{year}{1997};\hspace{0pt}\textbf{\bibinfo{volume}{119}}:\bibinfo{pag%
es}{233--247}.
\bibitem[{Loth and Dorgan(2009)}]{loth:2009}
\bibinfo{author}{Loth\xfnm[ E.]}, \bibinfo{author}{Dorgan\xfnm[ A.]}.
\newblock \bibinfo{title}{An equation of motion for particles of finite
  reynolds number and size}.
\newblock \emph{\bibinfo{journal}{Env Fluid Mech}}
  \bibinfo{year}{2009};\hspace{0pt}\textbf{\bibinfo{volume}{9}}:\bibinfo{pages%
}{187--206}.
\bibitem[{Calzavarini et~al.(2008{\natexlab{a}})Calzavarini, Kerscher, Lohse
  and Toschi}]{calzavarini:2008a}
\bibinfo{author}{Calzavarini\xfnm[ E.]}, \bibinfo{author}{Kerscher\xfnm[ M.]},
  \bibinfo{author}{Lohse\xfnm[ D.]}, \bibinfo{author}{Toschi\xfnm[ F.]}.
\newblock \bibinfo{title}{Dimensionality and morphology of particle and bubble
  clusters in turbulent flow}.
\newblock \emph{\bibinfo{journal}{J Fluid Mech}}
  \bibinfo{year}{2008}{\natexlab{a}};\hspace{0pt}\textbf{\bibinfo{volume}{607}%
}:\bibinfo{pages}{13--24}.
\bibitem[{Calzavarini et~al.(2008{\natexlab{b}})Calzavarini, Cencini, Lohse and
  Toschi}]{calzavarini:2008b}
\bibinfo{author}{Calzavarini\xfnm[ E.]}, \bibinfo{author}{Cencini\xfnm[ M.]},
  \bibinfo{author}{Lohse\xfnm[ D.]}, \bibinfo{author}{Toschi\xfnm[ F.]}.
\newblock \bibinfo{title}{Quantifying turbulence-induced segregation of
  inertial particles}.
\newblock \emph{\bibinfo{journal}{Phys Rev Lett}}
  \bibinfo{year}{2008}{\natexlab{b}};\hspace{0pt}\textbf{\bibinfo{volume}{101}%
}:\bibinfo{pages}{084504}.
\bibitem[{Volk et~al.(2008)Volk, Calzavarini, Verhille, Lohse, Mordant, Pinton
  et~al.}]{volk:2008b}
\bibinfo{author}{Volk\xfnm[ R.]}, \bibinfo{author}{Calzavarini\xfnm[ E.]},
  \bibinfo{author}{Verhille\xfnm[ G.]}, \bibinfo{author}{Lohse\xfnm[ D.]},
  \bibinfo{author}{Mordant\xfnm[ N.]}, \bibinfo{author}{Pinton\xfnm[ J.F.]},
  et~al.
\newblock \bibinfo{title}{Acceleration of heavy and light particles in
  turbulence: comparison between experiments and direct numerical simulations}.
\newblock \emph{\bibinfo{journal}{Physica D}}
  \bibinfo{year}{2008};\hspace{0pt}\textbf{\bibinfo{volume}{237}}(\bibinfo{num%
ber}{14-17}):\bibinfo{pages}{2084--2089}.
\bibitem[{Calzavarini et~al.(2009)Calzavarini, Volk, Bourgoin, L\'ev\^eque,
  Pinton and Toschi}]{calzavarini:2009}
\bibinfo{author}{Calzavarini\xfnm[ E.]}, \bibinfo{author}{Volk\xfnm[ R.]},
  \bibinfo{author}{Bourgoin\xfnm[ M.]}, \bibinfo{author}{L\'ev\^eque\xfnm[
  E.]}, \bibinfo{author}{Pinton\xfnm[ J.F.]}, \bibinfo{author}{Toschi\xfnm[
  F.]}.
\newblock \bibinfo{title}{Acceleration statistics of finite-sized particles in
  turbulent flow: the role of faxŽn forces}.
\newblock \emph{\bibinfo{journal}{J Fluid Mech}}
  \bibinfo{year}{2009};\hspace{0pt}.
\bibitem[{Qureshi et~al.(2007)Qureshi, Bourgoin, Baudet, Cartellier and
  Gagne}]{qureshi:2007}
\bibinfo{author}{Qureshi\xfnm[ N.M.]}, \bibinfo{author}{Bourgoin\xfnm[ M.]},
  \bibinfo{author}{Baudet\xfnm[ C.]}, \bibinfo{author}{Cartellier\xfnm[ A.]},
  \bibinfo{author}{Gagne\xfnm[ Y.]}.
\newblock \bibinfo{title}{Turbulent transport of material particles: An
  experimental study of finite size effects}.
\newblock \emph{\bibinfo{journal}{Phys Rev Lett}}
  \bibinfo{year}{2007};\hspace{0pt}\textbf{\bibinfo{volume}{99}}(\bibinfo{numb%
er}{18}):\bibinfo{pages}{184502}.
\bibitem[{Brown et~al.(2009)Brown, Warhaft and Voth}]{brown:2009}
\bibinfo{author}{Brown\xfnm[ R.D.]}, \bibinfo{author}{Warhaft\xfnm[ Z.]},
  \bibinfo{author}{Voth\xfnm[ G.A.]}.
\newblock \bibinfo{title}{Acceleration statistics of neutrally buoyant
  spherical particles in intense turbulence}.
\newblock \emph{\bibinfo{journal}{Phys Rev Lett}}
  \bibinfo{year}{2009};\hspace{0pt}\textbf{\bibinfo{volume}{103}}(\bibinfo{num%
ber}{19}):\bibinfo{pages}{194501}.
\bibitem[{Volk et~al.(2010)Volk, Calzavarini, L\'ev\^eque and
  Pinton}]{volk:2010}
\bibinfo{author}{Volk\xfnm[ R.]}, \bibinfo{author}{Calzavarini\xfnm[ E.]},
  \bibinfo{author}{L\'ev\^eque\xfnm[ E.]}, \bibinfo{author}{Pinton\xfnm[]}.
\newblock \bibinfo{title}{Dynamics of inertial particles in a turbulent von
  karman flow}.
\newblock \emph{\bibinfo{journal}{arXiv:10014943v1}}
  \bibinfo{year}{2010};\hspace{0pt}.
\bibitem[{Homann and Bec(2010)}]{homann:2010}
\bibinfo{author}{Homann\xfnm[ H.]}, \bibinfo{author}{Bec\xfnm[ J.]}.
\newblock \bibinfo{title}{Finite-size effects in the dynamics of neutrally
  buoyant particles in turbulent flow}.
\newblock \emph{\bibinfo{journal}{J Fluid Mech}}
  \bibinfo{year}{2010};\hspace{0pt}\textbf{\bibinfo{volume}{651}}:\bibinfo{pag%
es}{81--91}.
\bibitem[{Schiller and Naumann(1933)}]{schiller:1933}
\bibinfo{author}{Schiller\xfnm[ L.]}, \bibinfo{author}{Naumann\xfnm[ A.]}.
\newblock \bibinfo{title}{†ber die grundlegenden berechungen bei der
  schwerkraftaufbereitung}.
\newblock \emph{\bibinfo{journal}{Vereines Deutscher Ingenieure}}
  \bibinfo{year}{1933};\hspace{0pt}\textbf{\bibinfo{volume}{77}}:\bibinfo{page%
s}{318--320}.
\bibitem[{Clift et~al.(1978)Clift, Grace and Weber}]{clift:1978}
\bibinfo{author}{Clift\xfnm[ R.]}, \bibinfo{author}{Grace\xfnm[ J.R.]},
  \bibinfo{author}{Weber\xfnm[ M.E.]}.
\newblock \emph{\bibinfo{title}{Bubbles, Drops and Particles}}.
\newblock \bibinfo{publisher}{Academic Press (1978), Dover (2005)};
  \bibinfo{year}{1978}.
\bibitem[{Burton and Eaton(2005)}]{burton:2005}
\bibinfo{author}{Burton\xfnm[ T.M.]}, \bibinfo{author}{Eaton\xfnm[ J.K.]}.
\newblock \bibinfo{title}{Fully resolved simulations of particle-turbulence
  interaction}.
\newblock \emph{\bibinfo{journal}{J Fluid Mech}}
  \bibinfo{year}{2005};\hspace{0pt}\textbf{\bibinfo{volume}{545}}:\bibinfo{pag%
es}{67--111}.
\bibitem[{Elghobashi and Truesdell(1992)}]{elghobashi:1992}
\bibinfo{author}{Elghobashi\xfnm[ S.]}, \bibinfo{author}{Truesdell\xfnm[
  G.C.]}.
\newblock \bibinfo{title}{Direct simulation of particle dispersion in a
  decaying isotropic turbulence}.
\newblock \emph{\bibinfo{journal}{J Fluid Mech}}
  \bibinfo{year}{1992};\hspace{0pt}\textbf{\bibinfo{volume}{242}}:\bibinfo{pag%
es}{655--700}.
\bibitem[{R. and R.J.(1992)}]{mei:1992}
\bibinfo{author}{R.\xfnm[ M.]}, \bibinfo{author}{R.J.\xfnm[ A.]}.
\newblock \bibinfo{title}{Flow past a sphere with an oscillation in the
  free-stream and unsteady drag at finite reynolds number}.
\newblock \emph{\bibinfo{journal}{J Fluid Mech}}
  \bibinfo{year}{1992};\hspace{0pt}\textbf{\bibinfo{volume}{237}}:\bibinfo{pag%
es}{323--341}.
\bibitem[{Mordant and Pinton(2000)}]{mordant:2000}
\bibinfo{author}{Mordant\xfnm[ N.]}, \bibinfo{author}{Pinton\xfnm[ J.F.]}.
\newblock \bibinfo{title}{Velocity measurements of a sphere settling in a fluid
  at rest}.
\newblock \emph{\bibinfo{journal}{Eur Phys J B}}
  \bibinfo{year}{2000};\hspace{0pt}\textbf{\bibinfo{volume}{18}}:\bibinfo{page%
s}{343--352}.
\bibitem[{Dorgan and Loth(2007)}]{dorgan:2007}
\bibinfo{author}{Dorgan\xfnm[ A.J.]}, \bibinfo{author}{Loth\xfnm[ E.]}.
\newblock \bibinfo{title}{Efficient calculation of the history force at finite
  reynolds numbers}.
\newblock \emph{\bibinfo{journal}{Int J Multiphase Flow}}
  \bibinfo{year}{2007};\hspace{0pt}\textbf{\bibinfo{volume}{33}}:\bibinfo{page%
s}{833--848}.
\bibitem[{van Hinsberg et~al.(2010)van Hinsberg, ten Thije~Boonkkamp and
  Clercx}]{hinsberg:2010}
\bibinfo{author}{van Hinsberg\xfnm[ M.]}, \bibinfo{author}{ten
  Thije~Boonkkamp\xfnm[ J.]}, \bibinfo{author}{Clercx\xfnm[ H.]}.
\newblock \bibinfo{title}{An efficient, second order method for the
  approximation of the basset history force}.
\newblock \emph{\bibinfo{journal}{arXiv:10080833v1}}
  \bibinfo{year}{2010};\hspace{0pt}.

\end{thebibliography}
\end{document}